%% file: main.tex
\documentclass[sigconf]{acmart}
\usepackage{graphicx} 
\usepackage{array} 
\usepackage[most]{tcolorbox}

\usepackage{algorithmic}
\usepackage{textcomp}
\usepackage{xcolor}
\usepackage{booktabs}
\usepackage{float}
\usepackage{tabularx}
\usepackage{caption}
\usepackage{comment}
\usepackage{enumitem}
\usepackage{subcaption}
\usepackage{multirow}

\acmConference[SIGCITE '25]{Proceedings of the 26th ACM Annual Conference on Cybersecurity and Information Technology Education (ACM SIGCITE 2025)}{November 6--8, 2025}{Sacramento, California, USA}
\acmYear{2025}
\copyrightyear{2025}
\acmISBN{978-1-4503-XXXX-X/25/11}   
\acmDOI{10.1145/nnnnnnn.nnnnnnn}     
\setcopyright{none}

\title{Enhancing Software Testing Education: \\ Understanding Where Students Struggle}

\author{Shiza Andleeb}
\affiliation{%
  \institution{The University of Alabama}
  \city{Tuscaloosa}
  \country{USA}
}
\email{sandleeb@crimson.ua.edu}

\author{Teo Mendoza}
\affiliation{%
 \institution{Willamette University}
  \city{Salem}
  \country{USA}
}
\email{tjmendoza@willamette.edu}

\author{Lucas Cordova}
\affiliation{%
  \institution{Willamette University}
  \city{Salem}
  \country{USA}
}
\email{lpcordova@willamette.edu}

\author{Gursimran Walia}
\affiliation{%
  \institution{Augusta University}
  \city{Atlanta}
  \country{USA}
}
\email{gwalia@augusta.edu}

\author{Jeffrey Carver}
\affiliation{%
  \institution{The University of Alabama}
  \city{Tuscaloosa}
  \country{USA}
}
\email{carver@cs.ua.edu}



\begin{document}

\begin{abstract}
Effective software testing is critical for producing reliable and secure software, yet many computer science students struggle to master the foundational concepts required to construct comprehensive test suites. While automated feedback tools are widely used to support student learning, it remains unclear which testing concepts are most frequently misunderstood and how these misunderstandings are reflected in students’ test suite revisions.

This study examines the specific testing concepts that lead students to make ineffective changes—those that fail to improve code coverage—during test suite development. Leveraging an automated feedback tool in a senior-level software testing course, we analyzed student submissions from two assignments to identify prevalent conceptual gaps and patterns of unproductive modification. Our results reveal that decision coverage and exception handling are persistent challenges, and that students most often make superficial or method-level changes that do not enhance coverage.

These findings provide actionable insights for educators, researchers, and tool designers. By pinpointing the concepts that most often contribute to poor testing outcomes, we can refine feedback systems, target instruction to address persistent misconceptions, and more effectively support students in developing robust, maintainable test suites.

\end{abstract}

\maketitle

\section{Introduction}
\input{sec_Introduction}
\section{Related Work}
\input{sec_RelatedWork}
\section{Study Design}
\input{sec_StudyDesign}
\section{Data Analysis and Discussion}
\input{sec_DataAnalysis_and_Discussion}

\section{Threats to Validity}
\input{sec_Threats}
\section{Conclusion}
\input{sec_Conclusion}

\bibliographystyle{ACM-Reference-Format}
\bibliography{references}
\end{document}

%% file: sec_Introduction.tex
Software systems rely on the integration of numerous components, where each component's reliability critically affects the entire system's stability~\cite{chen2018microservices, zhou2019latent}. 
Recent empirical studies demonstrate that inadequate unit testing remains a primary cause of production failures, with Zhang et al. finding that 58\% of catastrophic failures in distributed systems could have been prevented by effective unit tests~\cite{zhang2014simple}. 
This challenge is amplified in microservice architecture systems, where a single poorly-tested service can trigger cascading failures across the entire system~\cite{toffetti2015architecture, wang2018runtime}. 
The economic impact is substantial: poor testing costs the US economy \$60-70 billion annually~\cite{charette2005software}. 

Yet educational research reveals that many computer science students struggle to grasp core software testing concepts, resulting in incomplete or ineffective test suites~\cite{edwards2014difficulty, buffardi2015machine}, potentially introducing weaknesses that propagate through integrated systems. 
This challenge is compounded by the lack of clarity among educators regarding which specific concepts present the greatest obstacles for learners~\cite{messer2024systematic}.

Recent research underscores the importance of conceptual feedback in software testing education. Studies have shown that providing students with feedback on missing fundamental testing concepts—such as branch coverage, equivalence class partitioning, and exception handling—leads to greater improvements in test suite quality than simply identifying missing test cases~\cite{kuechler2020inquiry, fan2017conceptgrader}. 
Despite these advances, students frequently make changes to their test suites that fail to improve code coverage, indicating persistent misunderstandings of foundational concepts such as decision coverage, boundary value analysis, and robust exception testing~\cite{more2025efficient, keuning2023mapping}.

The complexity of modern software systems has heightened the need for robust testing practices, making it essential for students to develop a deep, conceptual understanding of testing principles. 
Effective testing requires more than just writing test cases—it involves identifying edge cases, handling exceptions, and achieving comprehensive coverage of code paths~\cite{clarke2014integrating, roychoudhury2022testsmells}. 
However, there remains limited research on which testing concepts are most difficult for students, and how these difficulties manifest in their test suite modifications.

Building on prior work, this study aims to bridge the gap between conceptual understanding and practical application in software testing education. 
Specifically, we focus on identifying which testing concepts are most frequently misunderstood and how these misunderstandings lead to ineffective changes—defined as modifications to a test suite that can leave source code untested and vulnerable. 
By examining student test suite submissions and their responses to automated, concept-level feedback, we seek to uncover persistent patterns of struggle and provide actionable insights for improving feedback mechanisms and instructional strategies.

To drive this study, we pose a set of research questions.
First, we aim to discover whether ineffective changes, i.e., those changes to student test suites that do not increase test coverage, are related to misunderstandings of specific testing concepts.

This goal leads to the first research question.
\begin{tcolorbox}[enhanced,drop shadow]
\textbf{RQ1: Which fundamental software testing concepts are most frequently associated with students making ineffective changes to their test suites?}
\end{tcolorbox}

Second, we aim to identify types of ineffective changes.
This investigation involves identifying actions, such as method changes that do not enhance coverage, and categorizing different types of ineffective attempts. 
This goal leads to the second research question.

\begin{tcolorbox}[enhanced,drop shadow]
\textbf{RQ2 : What types of ineffective modifications do students commonly make to their test suites, and why do these changes fail to improve test coverage?}
\end{tcolorbox}

By analyzing patterns in student submissions, we identify conceptual gaps and modification types that signal misunderstanding, providing insights for educators, tool designers, and researchers to refine automated feedback, develop targeted materials, and build effective interventions~\cite{messer2024systematic, zhang2025autograder, liu2024stepgrade}. 

%% file: sec_RelatedWork.tex
Software testing pedagogy has evolved from automated tools (e.g., WebCAT, Marmoset) 
to data analysis of how students reflect on their testing practices and understand core testing concepts in a manner that goes beyond a trial-and-error-based development approach \cite{wong2011}, ~\cite{garousi2020software}, ~\cite{aniche2019pragmatic}. 

This study builds upon existing research on test suite feedback to identify why students struggle to achieve full code coverage despite receiving conceptual guidance. 
Previous research has shown that sharing coverage metrics motivates students to exercise more code paths, but without focusing on the conceptual understanding of core testing principles. 
The early work of [anonymized] addressed this limitation by comparing detailed (line-level) and conceptual (concept-level) feedback, demonstrating that concept-level feedback yielded higher-quality test suites and code security. 
However, students still struggled to achieve full coverage even with conceptual feedback. 
This work extends those earlier findings by examining submission-level differences to identify and characterize ineffective revisions to test suites. 
We focus specifically on the underlying concepts that students struggle to implement when revising their test suites.

Prior work has also focused on performance outcomes (e.g., coverage statistics) and self-reported perceptions~\cite{delgado-perez2024mutation} of student testers.
However, we are not aware of any work that has attempted to analyze ineffective change patterns tied to conceptual misunderstandings. 
By closing this gap, this work can offer recommendations to the feedback systems (like the one used by [anonymized]) to adapt to these patterns. 

Pedagogical interventions (e.g., Test-Driven Development (TDD)~\cite{edwards2004tdd}) have highlighted the improvement of student understanding of testing strategies by focusing on outcomes or final scores.
While this practice (developing tests before source code) encourages students to adopt a more reflective approach, it does not reveal which testing concepts cause repeated but ineffective edits to the test suite~\cite {Schon1983}.

Web-based testing frameworks~\cite{clarke2010automated, clarke2012empirical} and gamified testing environments~\cite{lemos2020active} enhance engagement, but are not able to point out fundamental testing concepts mapped to the submitted work (codebase) that can inform the restricted design and feedback mechanism. This paper tries to close that gap and support testing pedagogy. 

Several researchers have emphasized the benefits of integrating testing tools into the curriculum, such as enhancing students’ ability to debug and refine their code through iterative feedback \cite{clarke2014integrating, lim2023pyrates}. Automated assessment environments and real-time feedback systems have been shown to improve student engagement, increase assignment completion rates, and foster the development of essential testing skills \cite{zhang2025autograder, messer2024systematic}. For example, Lim et al.\cite{lim2023pyrates}\ demonstrated that adaptive, game-based feedback can significantly improve novice programmers’ robustness testing skills, while Clarke et al.\cite{clarke2012empirical}\ reported that collaborative testing platforms support deeper understanding of software quality assurance processes.

However, these studies often do not address the areas where students struggle despite receiving automated feedback. This limitation hinders the identification of persistent misconceptions or gaps in understanding, which are critical for informing more targeted instructional approaches. Recent systematic reviews have called for research that goes beyond tool adoption to analyze student errors and misconceptions in depth, emphasizing the need for conceptual scaffolding and adaptive feedback mechanisms that respond to specific learning challenges \cite{messer2024systematic, keuning2023mapping}.

%% file: sec_StudyDesign.tex
This overviews the study setting, participants, study procedure, variables, data collection process, and ethical approval.

\subsection{Study Setting}
We conducted the study in a split-level Testing and Quality Assurance course at a large, public university in the southeastern United States. 
This elective course focused on the principles and best practices for creating reliable and maintainable software, with an emphasis on test design, test automation, and quality assurance processes.
The goal of the course was to prepare students for industry roles by providing hands-on experience with testing tools and techniques, fostering critical thinking, and encouraging a systematic approach to software quality.
The course covered essential software testing concepts, including validation, verification, defect management, white-box testing, and black-box testing. 
Students learned about coverage criteria, test automation, mutation testing, and how to write effective test plans. 
The course also included two hands-on assignments where students applied these concepts to real-world scenarios.
We used these two assignments for the study.

\subsection{Participants}
The study included a diverse mix of 46 domestic and international students, ensuring a broad representation of backgrounds and experiences in software testing. 

\subsection{Study Procedure}
To facilitate this study and gather data, we used the [anonymized] tool, a web-based automated feedback tool. 
This tool enhances student learning about software testing by offering tailored, focused guidance on creating effective test suites. 
The tool compares a student's test suite for a given piece of code against a reference test suite that fully tests that code.
The tool identifies mismatches between the student's test suite and the reference test suite.
It then matches the missing tests to the fundamental testing concepts they embody, such as boundary testing, exception handling, or input validation.
The tool then provides the student with feedback about which fundamental testing concepts are not fully exercised in their submitted test suite, along with guidance about how to address the deficiency.

This approach encourages the student to think critically and improve their test suite independently, rather than providing them with the exact problems in their test suite.

Before starting their assignments, a structured tutorial introduced the students to the automated feedback tool.
This tutorial included a YouTube video and detailed written instructions. 
The video demonstrated how to submit test suites, interpret feedback, and refine test cases based on the system’s automated responses. 
The students were required to watch the video and read the instructions before using the tool.
The students also had the contact details of the instructor and the tool developers if they had further questions or encountered difficulties while using the tool. 
This approach ensured that students had multiple avenues for support and could effectively engage with the feedback process.

Students used white-box testing in two assignments (described below).
For each assignment, they received code for a small program and had to create a test suite that fully tested that implementation.
The students were not allowed to modify the given code, so they focused on developing a test suite that achieved full coverage.

Each test suite submission triggered an automated analysis that provided students with two types of feedback:
\begin{itemize}
    \item \textbf{Conceptual Feedback} -- information about which fundamental testing concepts (e.g., \textit{boundary conditions} or \textit{input validation}) remain unexercised. 
    \item \textbf{Detailed Feedback} -- results from a code coverage tool indicating exactly which parts of the code the test suite did not fully test.
    It pinpoints untested methods and branches, clearly indicating areas of improvement and using visual cues to distinguish tested and untested paths.
\end{itemize}

Conceptual feedback is intentionally less explicit, encouraging students to develop a better understanding of the fundamental testing concepts and determine why the test suite is incomplete so they can make the necessary changes. 
    
As the students created their test suites, they could submit and resubmit them to the automated feedback tool, which provided information about their completeness.
We operated under the assumption that when making changes to a test suite, a student was trying to increase test coverage.
When successive submissions yielded no coverage gain, we inferred that the student's test strategy was ineffective.
Although instructions urged students to strive for the ``most complete, yet smallest'' test suite, we did not penalize them for submitting larger or redundant test suites; however, the tool flagged redundant test cases so we could analyze those patterns.

The first assignment was \textit{MyDateImpl}, a program that implemented an immutable date class.
Students wrote test cases for the class's input validation, edge cases, and date-related operations.
The second assignment was \textit{CircularQueue}, a program implementing a circular queue data structure.
Students wrote test cases involving queue operations, state management, and ensuring the correct behavior of the enqueue and dequeue operations under various conditions. 
During Assignment 1, students received only conceptual feedback.
Mid-course analysis revealed that some students got ``stuck'' while developing their test suites (i.e., they had numerous consecutive test suite submissions that did not increase coverage)
Therefore, for Assignment 2, we configured the tool to recognize when a student was stuck (based on submitting multiple consecutive test suite submissions with no improvement) and changed the feedback to detailed feedback.

To identify which testing concepts students struggled with, we mapped each method in the assignments to the corresponding concept that it was designed to assess.
For example, compareTo() involves logical decisions and requires tests that satisfy true and false branches; therefore, it maps to decision coverage.
Likewise, enqueue() involves state transitions and boundary conditions, mapping to equivalence class partitioning. 

\subsection{Independent and Dependent Variables}
This study has two independent variables: type of feedback (conceptual vs. detailed) and assignment (Assignment 1 vs. Assignment 2).
The study had the following dependent variables: individual test suite submissions, missed testing concepts (per tool feedback), ineffective changes (consecutive submissions with zero coverage gain), and total number of submissions per student. 

\subsection{Study Ethics}
The [anonymized university] Institutional Review Board (IRB) approved this study. 
Participation was voluntary and anonymous.
Regardless of whether students consented to allow us to use their data for research, they could still received the same type of feedback.
Author [anonymized] was the instructor of the host course. 
His role during the course was limited to delivering the course content.
To maintain objectivity, he did not participate in the consenting process or data collection and was unaware of which students participated in the study.
After the course concluded, he assisted with data analysis but still did not have access to identifying information about the participants. 
We organized the study this way to prevent any undue influence on the students.

%% file: sec_DataAnalysis_and_Discussion.tex
This section presents our analysis of student submissions, organized around the two research questions. 
By examining where students consistently made ineffective changes and misunderstood core testing concepts, we identify high-leverage areas for improving testing instruction.

\subsection{RQ1:  Which Fundamental Testing Concepts
Are Most Associated with Ineffective Test
Changes?}

To determine which testing concepts students most struggled to apply, we analyzed every submission whose revised test suite did not increase coverage.  
For each such submission, our tool compared the student’s tests against a reference suite and recorded which fundamental concepts remained unaddressed.  
We tracked six key concepts: decision coverage, branch coverage, exception handling, equivalence class partitioning, boundary value analysis, and functionality verification.  
Table~\ref{misunderstoodConcepts} displays the percentage of submissions in each assignment for which each concept was misunderstood.

\begin{table}[!htb]
    \centering
    \caption{Percentage Distribution of Misunderstood Testing Concepts for Assignment 1 and 2}
    \label{misunderstoodConcepts}
    \input{tab_MisunderstoodConcepts}

\end{table}

Our analysis reveals that \textbf{decision coverage} and \textbf{branch coverage} were the most frequently missed concepts, together accounting for roughly half of all misunderstandings in both assignments.  This indicates a pervasive difficulty in constructing tests that exercise all control paths and logical branches.  \textbf{Exception handling} was also a major challenge: over one-third of students in Assignment 1 and an even larger fraction in Assignment 2 failed to include tests for invalid inputs or error conditions, suggesting that robustness considerations grow more challenging as assignments progress.  

While \textbf{equivalence class partitioning} was rarely overlooked in the first assignment, it became one of the top three misunderstood techniques in the second, highlighting a shift in where students focus their test‐design effort over time.  
Similarly, \textbf{boundary value analysis}, the practice of testing inputs at the edges of valid ranges, was omitted by a notable subset of students in both assignments, underscoring a consistent need for better instruction on edge‐case identification.  

In contrast, \textbf{functionality verification}—the most fundamental task of checking expected outputs for valid inputs—was misunderstood by only a small fraction of students, reflecting general competence with basic behavioral validation.  Taken together, these findings pinpoint the areas where students require the most pedagogical support and set the stage for our detailed examination of ineffective test‐suite modifications in RQ2.

\subsection{RQ2: What Types of Ineffective Test Suite Changes Do Students Make, and Why Do They Fail?}

Building upon the behavioral patterns identified in RQ1, this section presents a comprehensive analysis of specific ineffective testing modifications across two programming assignments. 

To examine the ineffective changes between consecutive test suites, we performed a detailed analysis using the \textbf{Python difflib} library. 
We used the tools from this library to compare sequences of code execution between two submissions. 
Specifically, we used the \textbf{difflib.ndiff()} function because it produces a readable output showing how a sequence of lines differs.

\begin{table}[!ht]
\centering
\caption{Type of Ineffective Changes}
\label{tab:ineffectivechanges}
\input{tab_IneffectiveChanges}

\end{table}

After extracting the differences in code, we organized them into the categories of ineffective changes shown in Table~\ref{tab:ineffectivechanges}.
We chose the categories based on their relevance to the common testing challenges seen in students' submissions and their impact on code coverage. 
For example, method modifications often indicate efforts to organize test cases.
However, if not done carefully, these changes may not improve coverage.
This classification helps us concentrate on areas where students require extra support or guidance.

We identified three categories of changes.
We placed each submission into a single category based on its main change type, making the categories mutually exclusive.
When a submission included multiple changes (such as modifications to methods and assert statements), we categorized it based on the most significant modification, typically the one that has the greatest impact on the test structure.
The \textbf{No Difference} category describes the situation where consecutive submissions had no differences in the submitted test code.
The \textbf{100\% Different} category describes the situation where the entire test suite was rewritten without any overlap with the previous submission.
The \textbf{Methods} category describes the situation when students add, remove, or change a test method (e.g., edit the code in a test method or edit the parameters, except assert statements, which are handled in the Assert Statements category) between consecutive test suites. 
The \textbf{Assert Statements} category describes the situation when a student adds or removes assert statements within their test methods. 
The \textbf{Other} category describes miscellaneous changes that do not fit the other categories, such as other structural modifications (i.e., simply moving code around, refactoring the test code, configuration changes, etc).

We further analyzed the categories of the ineffective changes. Table~\ref{tab:categories} presents the detailed categorization of these ineffective changes, revealing seven distinct patterns of problematic behavior. First, \textbf{superficial modifications}—such as trivial or redundant tests, single-method "testAll" approaches, reused inputs, cosmetic renaming, and empty placeholders—dominates student behavior.  Next, \textbf{assertion misuse} emerges frequently, with students relying on generic or incorrect assertions rather than validating specific behavior.  \textbf{Insufficient method coverage} is another common issue, as students often omit key methods, skip equality checks, or over-test a single method at the expense of others.  Students also show difficulty with \textbf{input partitioning}, failing to organize test cases into meaningful categories, and they tend to \textbf{overfocus on valid inputs}, neglecting error or exception scenarios.  A \textbf{happy-path bias} further illustrates their reluctance to test failure conditions.  Finally, an \textbf{over-reliance on exception testing alone} appears only occasionally.  Together, these patterns reveal that students predominantly make surface-level changes and misuse assertions, while deeper test-design challenges—such as comprehensive coverage and systematic partitioning—persist throughout their revisions.

\begin{table*}[!htb]
\centering
\caption{Common Ineffective Testing Modifications Across Two Assignments}
\input{tab_common_themes_misconceptions_2}
\end{table*}

\subsection{Discussion}

In this study, we have characterized pervasive patterns of ineffective test‐suite modifications and fundamental misconceptions in student testing practices. Our findings—from both research questions—converge on three core themes: surface‐level test changes, conceptual gaps in coverage criteria, and risk‐averse test design.

First, the dominance of \textbf{superficial modifications} (e.g., renaming tests, trivial test additions) and \textbf{assertion misuse} (e.g., \texttt{assertTrue(true)}) demonstrates that students often treat testing as a procedural checklist rather than an opportunity to specify and validate program behavior. This aligns with prior work showing that novice programmers focus on surface‐level changes when responding to feedback, neglecting deeper conceptual goals \cite{roychoudhury2022testsmells, kuechler2020inquiry}.

Second, misconceptions in \textbf{decision coverage} and \textbf{branch coverage} emerged as the most frequently misunderstood concepts, corroborating the literature’s identification of coverage criteria as challenging for learners \cite{keuning2023mapping}. Students’ difficulty with \textbf{boundary value analysis} and \textbf{equivalence partitioning} further indicates a lack of systematic test‐design strategies, reflecting earlier calls for stronger emphasis on partitioning techniques in curricula \cite{more2025efficient, clarke2014integrating}.

Third, the prevalence of \textbf{happy‐path bias} and inadequate \textbf{exception testing} suggests that students are reluctant to engage with error and edge‐case scenarios, undermining software robustness. This risk‐averse testing behavior echoes findings in industry‐focused studies, where practitioners similarly under‐test failure paths without explicit pedagogical support \cite{messer2024systematic}.

Taken together, these results underscore the need for testing education that integrates:
\begin{itemize}
  \item \textbf{Conceptual scaffolding} of coverage criteria and partitioning methods, rather than mere tool usage \cite{keuning2023mapping,kuechler2020inquiry}.
  \item \textbf{Structured test‐design frameworks} (e.g., requirement‐based planning, risk‐based prioritization) to guide systematic coverage and edge‐case identification \cite{clarke2014integrating, more2025efficient}.
  \item \textbf{Actionable, concept‐level feedback} that flags ineffective patterns (e.g., missing branches, redundant tests) and explicates their pedagogical rationale \cite{messer2024systematic, zhang2025autograder}.
  \item \textbf{Metacognitive reflection} activities (e.g., test‐planning worksheets, peer reviews) to promote self‐awareness of testing goals and iterative improvement \cite{liu2024stepgrade}.
\end{itemize}

The persistence of core misconceptions across assignments underscores the need for repeated, scaffolded practice and adaptive remediation. 
Future work should develop and evaluate \textbf{real‐time adaptive feedback systems} that dynamically identify student errors and deliver context‐aware guidance, and longitudinal studies to measure conceptual growth over a course or degree program.

By fostering deep \textbf{conceptual understanding} and \textbf{systematic thinking} in software testing, educators can help students move beyond procedural compliance toward robust test‐suite development, preparing them for the complexities of industrial practice.

%% file: tab_MisunderstoodConcepts.tex

    \begin{tabular}{lrr}
        \toprule
        \textbf{Test Concept} & \textbf{Assignment 1} & \textbf{Assignment 2} \\
        \midrule
        Decision Coverage & 89.7\% & 88.4\% \\
        Boundary Value Analysis & 12\% & 11.8\%\\
        Exceptions & 35.5\% & 42\% \\
        Equivalence & 3\% & 89.7\% \\
        Branch Coverage & 91\% & 96\% \\
        Functionality & 5.6\% & 3.9\% \\
        \bottomrule
    \end{tabular}


%% file: tab_IneffectiveChanges.tex
    \begin{tabular}{lrr}
        \toprule
        \textbf{Ineffective Changes} & \textbf{Assignment 1} & \textbf{Assignment 2} \\
        \midrule
        Methods & 38\% & 57\% \\
        Assert Statements & 33\% & 16\% \\
        Other & 20\% & 13\% \\
        No Difference & 8\% & 12\% \\
        100\% Different & 1\% & 2\% \\
        \bottomrule
    \end{tabular}


%% file: tab_common_themes_misconceptions_2.tex
\begin{tabular}{|p{.15\linewidth}|p{.61\linewidth}|p{.07\linewidth}|p{.07\linewidth}|}
\hline
\textbf{Category} & \textbf{Ineffective Modification} & \textbf{Assign. 1 Count} & \textbf{Assign. 2 Count} \\
\hline
\textbf{Superficial \newline Modifications} & Wrote redundant or trivial tests (added minimal logic or copied existing tests without meaningful variation) & 147 & 93 \\
\cline{2-4}
 & Wrote all logic in a single test method (used a testAll style without modularity or clarity) & 34 & 25 \\
\cline{2-4}
 & Reused identical inputs (repeated the same data without partitioning) & 25 & 14 \\
\cline{2-4} 
 & Renamed tests without improving logic (renamed test methods or inputs without altering test behavior or adding coverage) & 641 & 133 \\
\cline{2-4} 
 & Submitted empty or placeholder tests (created files with little or no actual testing logic) & 15 & 6 \\
\hline
\textbf{Overfocus on Valid Inputs} & Focused only on valid inputs (did not test invalid or error-triggering inputs) & 189 & 162 \\
\cline{2-4}
 & Reused passing values in tests (lacked diverse or unexpected test scenarios) & 166 & 82 \\
\hline
\textbf{Insufficient Method \newline Coverage} & Ignored required methods (left out important methods like isLeapYear or resizing behavior) & 367 & 74 \\
\cline{2-4}
 & Skipped equals/hashCode validation (avoided object comparison methods) & 64 & 36  \\
 \cline{2-4}
 & Focused heavily on a single method (repeatedly tested one method while neglecting others) & 78 & 26 \\
\hline
\textbf{Assertion Misuse} 
& Used assertTrue(true) or assertFalse(true) (overused generic assertions that do not validate specific behavior) & 272 & 41 \\
\cline{2-4}
& Misused or omitted assertions (used wrong assertions like assertNull instead of assertEquals) & 184 & 162 \\
\cline{2-4}
& Used if-else or print instead of assertions (printed outputs or manually inspected instead of asserting behavior) & 107 & 10 \\
\hline
\textbf{Lack of \newline Partitioning} 
& Repeated tests with similar inputs (didn't partition inputs across categories) & 254 & 137 \\
\cline{2-4}
& Missed edge case scenarios (failed to include edge conditions like boundary dates or wrap-around) & 148 & 106 \\
\hline
\textbf{Happy Path Bias} 
& Avoided failure and exception paths (skipped tests that would provoke exceptions or error-handling code) & 280 & 150 \\
\cline{2-4}
& Ignored system state verification (didn't examine internal conditions after operations) & 56 & 37 \\
\hline
\textbf{Over-reliance on Exception Testing} & Wrote only failure-based tests (used assertThrows without validating regular behavior) & 7 & 4 \\
\hline
\end{tabular}
\label{tab:categories}

%% file: sec_Threats.tex
We organize this section around the common types of validity threats.

\subsection{Internal Validity}
Internal validity threats impact the overall findings of the research.
The primary internal threat is that students could submit the same test suite code repeatedly with no limit. 
Because we did not limit the number of test suite submissions, the process may not have encouraged real iterative improvement but rather trial-and-error.
However, the results still provide interesting insights into what students thought would improve their test suites even if they did not seek to understand the underlying concepts fully.

\subsection{Construct Validity}
Construct validity threats impact the mapping of the data collected to the underlying theoretical construct.
The main threat in our study is that reported measures (feedback and coverage percentage) might not reference the true underlying concepts of student learning in software testing.

\subsection{External Validity}
External validity threats impact the extent to which the results can be generalized.
First, the students who participated came from one course at a single university.
Therefore, they may not be representative of all students. 
Second, the students wrote test cases for two specific programs.
If we used different programs, it is possible that the results would change.
To address both of these threats, we need to repeat the study in different locations with different programs.

%% file: sec_Conclusion.tex
The systematic analysis of student test‐suite modifications across two programming assignments demonstrates that elevating students’ testing proficiency requires pedagogical strategies beyond introducing tools and basic techniques. 
Our findings indicate that novices often fall back on superficial modifications, misuse assertions, and exhibit a ``happy‐path'' bias, revealing deep‐seated misconceptions about test design. 
These findings extend beyond software testing to illuminate broader challenges in computing education: students often respond to feedback with surface-level changes rather than addressing fundamental conceptual gaps. 

Our findings offer immediate, actionable guidance for computing educators. To transform student testing practices from superficial modifications to systematic improvement, we recommend:

\begin{itemize}
  \item \textbf{Core Testing Principles:}  Emphasize the rationale behind coverage criteria (e.g., decision and branch coverage), boundary value analysis, equivalence partitioning, and exception testing.  Explicitly teach how each principle contributes to fault detection and software robustness \cite{keuning2023mapping, kuechler2020inquiry}.
  \item \textbf{Systematic Test‐Design Frameworks:}  Introduce structured methodologies, e.g., partitioning strategies, test planning, and risk‐based test prioritization, to guide students from specifications to comprehensive test suites~\cite{clarke2014integrating, more2025efficient}.
  \item \textbf{Rich, Conceptual Feedback Mechanisms:}  Leverage automated tools or instructor annotations that identify ineffective patterns (e.g., redundant tests, missing branches) and provide actionable, concept‐level guidance, rather than solely reporting coverage metrics \cite{messer2024systematic, zhang2025autograder}.
  \item \textbf{Metacognitive Scaffolding:}  Encourage reflective practices, such as test‐planning worksheets or peer‐review sessions, that prompt learners to articulate testing goals, evaluate coverage choices, and iteratively refine their approach~\cite{liu2024stepgrade}.
\end{itemize} 

The persistence of certain ineffective behaviors across assignments highlights the need for repeated, scaffolded practice and targeted remediation. 
Future work should explore adaptive feedback systems that dynamically surface common misconceptions in real time and integrate longitudinal assessments to measure conceptual growth. 
We encourage the computing education community to replicate this analysis.
What ineffective modification patterns emerge when students struggle with system integration testing, security vulnerabilities, or machine learning concepts? 

By grounding curricula in these pillars—principles, frameworks, feedback, and reflection—educators can move students beyond procedural ``checklist'' approaches not just in testing, but across computing. 
This shift from surface-level quality to deeper conceptual understanding is essential for preparing students to tackle the complex, interconnected systems that define modern computing. 